\documentclass[%
reprint,
twocolumns,
aps,
prl,
showpacs,
notitlepage,
nobibnotes,
superscriptaddress
]{revtex4-2}

\usepackage{float}
\usepackage{bigints}
\usepackage{psfrag}
\usepackage{grffile}
\usepackage{verbatim}
\usepackage{microtype}
\usepackage{multirow}
\usepackage{enumitem}
\usepackage{amsmath}
\usepackage{amssymb}
\usepackage{amsthm}
\usepackage{mathrsfs}
\usepackage{mathtools}
\usepackage{graphicx}
\usepackage{bbm}
\usepackage{subfig}
\usepackage[linesnumbered,ruled]{algorithm2e}
\usepackage{color}
\usepackage{tcolorbox}

\usepackage{ragged2e}

\definecolor{myblue}{rgb}{0.153,0.322,0.706}
\usepackage[colorlinks,linkcolor=myblue,urlcolor=myblue,citecolor=myblue]{hyperref}
\usepackage{geometry}
\usepackage{url}
\usepackage{hyperref}
\usepackage{xcolor}
\usepackage{soul}
\newcommand{\be}{\begin{equation}}
\newcommand{\ee}{\end{equation}}

\def\bc{\begin{center}}
\def\ec{\end{center}}
\def\bea{\begin{eqnarray}}
\def\eea{\end{eqnarray}}

\DeclareCaptionJustification{justified}{\justifying}
\definecolor{airforceblue}{rgb}{0.36, 0.54, 0.66}
\definecolor{brickred}{rgb}{0.8, 0.25, 0.33}
\definecolor{amber}{rgb}{1.0, 0.75, 0.0}
\definecolor{applegreen}{rgb}{0.55, 0.71, 0.0}
\definecolor{magenta}{rgb}{0.965, 0, 0.859}

\newcommand{\newtext}[1]{{\color{black}{#1}}}

\newcommand{\newtexttwo}[1]{{\color{black}{#1}}}

\begin{document}

\title{Dynamical fluctuations of random walks in higher-order networks}

\author{Leonardo Di Gaetano}
%\thanks{These two authors contributed equally}
%\email{di-gaetano\_leonardo@phd.ceu.edu}
\affiliation{Department of Network and Data Science, Central European University, 1100 Vienna, Austria}

\author{Giorgio Carugno}
%\email{giorgio.carugno@kcl.ac.uk}
\affiliation{Department of Mathematics, King’s College London, Strand, London WC2R 2LS, UK}

\author{Federico Battiston}
\email{battistonf@ceu.edu}
\affiliation{Department of Network and Data Science, Central European University, 1100 Vienna, Austria}

\author{Francesco Coghi
%\normalfont\textsuperscript{*,}
}
%\thanks{These two authors contributed equally}
\email{francesco.coghi@su.se}
\affiliation{Nordita, KTH Royal Institute of Technology and Stockholm University, Hannes Alfvéns väg 12, SE-106 91 Stockholm, Sweden}

\date{\today}

\begin{abstract}
Although higher-order interactions are known to affect the typical state of dynamical processes giving rise to new collective behavior, how they drive the emergence of rare events and fluctuations is still an open problem. We investigate how fluctuations of a dynamical quantity of a random walk exploring a higher-order network arise over time. \newtext{In the quenched case, where the hypergraph structure is fixed, through large deviation theory we show that the appearance of rare events is hampered in nodes with many higher-order interactions, and promoted elsewhere. 
Dynamical fluctuations are further boosted in an annealed scenario, where both the diffusion process and higher-order interactions evolve in time. Here, extreme fluctuations generated by optimal higher-order configurations can be predicted in the limit of a saddle-point approximation.}
Our study lays the groundwork for a wide and general theory of fluctuations and rare events in higher-order networks.

\end{abstract}

\maketitle

The appearance of fluctuations in dynamical processes is central in determining the future evolution of many real-world systems~\cite{Albeverio2006}. The emergence of rare events may be bolstered or hindered by the hosting complex environment, often conveniently modeled as a complex network~\cite{Barrat2008,Newman2010,Latora2017}. Large fluctuations in complex networks have been studied across a variety of processes, including percolation~\cite{Bianconi2017,Bianconi2018,Coghi2019,Kumar2020}, spreading~\cite{Hindes2016,Hindes2017}, and transport~\cite{Chen2014,Chen2015,Staffeldt2019,Gupta2021}. A stream of research has focused on random walks as a versatile model of diffusion in discrete spaces~\cite{Noh2004,Rosvall2005,Gomez-Gardenes2008,Burda2009,Sinatra2011} and on their rare event properties~\cite{Kishore2011,Kishore2012,Gandhi2022}. Large deviation theory has revealed that low-degree nodes are more susceptible than hubs to the appearance of atypical loads, possibly leading to dynamical phase transitions~\cite{DeBacco2016,Coghi2018,Gutierrez2021,carugno2023delocalization}.

%higher-order interactions
Despite their success, graphs can only provide a constrained description of real-world systems, as links are inherently limited to model pairwise interactions only~\cite{battiston2020networks, battiston2022higher,berge1973graphs}. Yet, from social~\cite{patania2017shape,benson2018simplicial,cencetti2021temporal,musciotto2022beyond} to biological~\cite{klamt2009hypergraphs,petri2014homological,giusti2016two,zimmer2016prediction} networks, in a wide variety of real-word systems interactions may occur among three or more units at a time. 
Interestingly, taking into account higher-order interactions has shown to lead to new collective phenomena in a variety of dynamical processes~\cite{battiston2021physics}, including diffusion~\cite{schaub2020random,Carletti2020b}, contagion~\cite{iacopini2019simplicial, neuhauser2020multibody,ferraz2023multistability}, synchronization~\cite{lucas2020multiorder, skardal2020higher, millan2020explosive, gambuzza2021stability, zhang2023higher}, percolation~\cite{di2024percolation} and evolutionary games~\cite{alvarez2021evolutionary,civilini2021evolutionary,civilini2024explosive}. While such studies have focused on characterising dynamical behavior at the typical state, understanding fluctuations and rare events %statistics 
driven by the presence of higher-order interactions is to this day still an open problem.

To this end, in this work we propose a study of fluctuations and rare events on higher-order networks using large-deviation theory tools. We focus on random walks on higher-order networks and on an 
%particular time-additive 
observable that monitors the time the random walker spends in certain regions of the hypergraph. Our study reveals how fluctuations arise in time for a random walk on a fixed hypergraph structure (\textit{quenched} case), and which higher-order structure is optimal to achieve them (\textit{annealed} case). In the quenched case the density of higher-order interactions regulates fluctuations of
occupation times, which are hampered around well-connected nodes and enhanced elsewhere.
\newtext{In the annealed case, where the structure of interactions is not \textit{a-priori} fixed, the random walk dynamics select the optimal higher-order structure that maximises fluctuations and rare events are boosted.
}

\newtexttwo{In the following, we present a computationally easy-to-handle hypergraph model to introduce a theory of fluctuations for higher-order networks.} Our theory and results are further validated in the Supplemental Material (SM) by means of extensive numerical simulations on a wide variety of more complex structures with local heterogeneity and with / without star-like structure, as well as more general dynamics of biased random walks.

\vspace{0.5cm}

\paragraph{Model }

\begin{figure}
    \centering
    \includegraphics[scale=0.6]{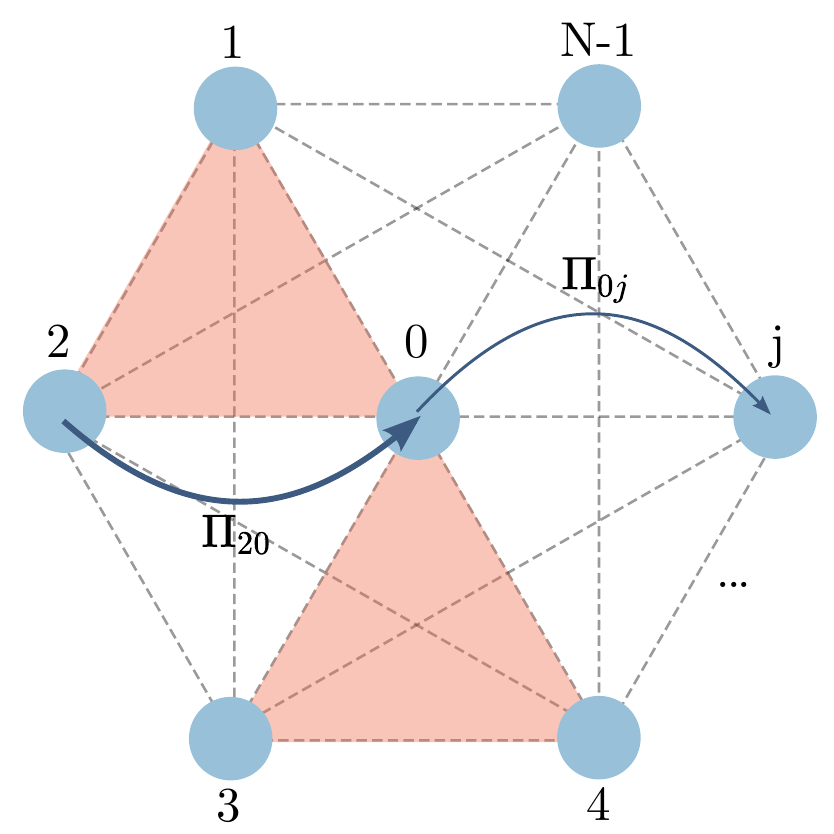}
    \caption{ Illustration of our model. 
    Dashed lines represent pairwise interactions that form the underlying complete graph. In pink, two higher-order interactions connect the core node $0$ with the peripheral nodes ($1,2$), and ($3,4$). The random walk's dynamics are represented by arrows departing from certain nodes and pointing towards others, where different thicknesses refer to different jump probabilities.
    }
    \label{fig0}
\end{figure}

We consider a hypergraph $G=(V,E)$, where $V$ represents the set of nodes, and $E= \{E_1,E_2,\dots,E_M \}$ the set of hyperedges, i.e., $E_m$ is an unordered collection of nodes belonging to the same hyperedge $m$. We focus in particular on an illustrative structure consisting of a \textit{core} node, labelled $0$, connected with \textit{peripheral} nodes through a varying number of higher-order connections, labelled by $i \in \{1, \dots, N-1\}$. As shown in Fig.~\ref{fig0}, the graph is composed by $|V|=N$ nodes, a fully connected pairwise structure, i.e. ${N \choose 2}$ binary edges $E_{i(2N-i-1)/2+j} = \{ i,j \}$ for $(i,j) \in [0,N-1]^2$ and $i < j$, and a %\st{binomially-distributed random} 
number \newtext{$\eta$ drawn from a binomial distribution} %\st{$H$} 
of parameter $p \in [0,1]$ of three-body interactions $E_{N(N-1)/2 + i} = \{ 0, i, j \}$ where $i$ is an odd node and $j-i=1$, i.e., all triangular interactions are centered in $0$. 
We constrain the higher-order structure so that each peripheral node can participate in at most one three-body interaction. As we will show, for this symmetric model, non-pairwise interactions affect the statistics of the core occupation time only through their total number $\eta$. In particular, the probability of drawing a hypergraph with a number of three-body interactions $H = \eta$ is given by
\begin{equation}\label{eq:probability_triangles}
    \mathbb{P}(\eta) \coloneqq \mathbb{P}(H=\eta) = {N_\triangle \choose \eta} p^\eta (1-p)^{N_\triangle - \eta} \quad ,
\end{equation}
where $N_\triangle = \text{ceil} \left[(N-2)/2\right]$ is the maximum number of possible three-body interactions that the hypergraph can have.

In summary, 
%for the model we consider here, 
$G$ is as an instance of an ensemble of hypergraphs whose higher-order structure is fully described by two parameters only, namely $N$ and $p$. 

%%Random walk on G and observable
We consider on $G$ an $n$-step discrete-time random walk $X = \{ X_l \}_{l=1}^n$, where $X_l$ denotes the node where the random walk sits at time $l$~\cite{Carletti2020b}. The random walk 
%is characterized by 
follows an unbiased dynamics given by the transition matrix $\Pi = \left\lbrace \pi_{ij} \right\rbrace$ whose entries are
\begin{equation}
    \label{eq:RWTransition}
    \pi_{ij} = \frac{k_{ij}^H}{\sum_{l=1}^N k_{il}^H} \ ,
\end{equation}
where $k_{ij}^H$ represents the hyperdegree, i.e., the number of nodes, excluding $i$, that are present in the hyperedges that are common to $i$ and $j$  
\newtexttwo{(see \textit{Appendix A} for details on how to derive the transition matrix).}
As the random walk explores the graph, it collects information in the form of the time-additive observable
\begin{equation}
    \label{eq:Observable}
    T_n = \frac{1}{n} \sum_{l=1}^n \delta_{X_l,0} \ ,
\end{equation}
which measures the fraction of time the random walk has spent on the core node $0$ up to time $n$. In the limit of $n\rightarrow \infty$, the typical fraction of time $T_{\eta,\text{typ}}$ the walker spends in $0$ for a number $H = \eta$ of three-body interactions reads~\cite{Carletti2020b}
\begin{equation}
T_{\eta,\text{typ}} = \frac{4 \eta + N -1}{8 \eta +(N-1)^2} \ .
\label{eq:TypicalTime}
\end{equation}
The higher the number of triangular interactions, the better connected the core with the periphery of the graph, and the longer the time the random walk will spend in $0$. 
Having delineated the typical behavior of the dynamical process, we now focus on its finite-time fluctuations. We consider dynamical fluctuations in two different physical scenarios. First, we study the mean behavior of rare events of $T_n$ over the ensemble of possible hypergraphs of our model (quenched case). Then, at the expense of an entropic cost associated with the logarithm of $P(\eta)$ in \eqref{eq:probability_triangles}, we let the random walk choose the optimal hypergraph that generates a particular atypical fluctuation of $T_n$ (annealed case). 
Results for more complex higher-order topologies, and for more general dynamics considering random walks biased on the higher-order structure, are qualitatively consistent and illustrated in the SM.

\vspace{0.5cm}

\paragraph{Quenched fluctuations}

In the quenched scenario, we consider averaged fluctuations in static hypergraph structures with $\eta$ three-body interactions and investigate how higher-order network configurations impact the dynamics of random walks. To do so, we employ large deviation theory~\cite{DenHollander2000,Touchette2009,Dembo2010}, making use of the leading scaling behavior of the probability distribution $\mathbb{P}_{\eta,n}(t) \coloneqq \mathbb{P}_{\eta,n}(T_n=t)$ that is exponential in time, i.e., 

\begin{equation}
    \label{eq:LDP}
    \mathbb{P}_{\eta,n}(t) = e^{-n I_\eta(t) + o(n)} \ ,
\end{equation}
where $I_\eta(t)$ is the non-negative large-deviation rate function containing the relevant information about rare events and $o(n)$ denotes sub-linear corrections in $n$. 
Evaluating $I_\eta$ directly is often non-trivial, thus we resort to a change of ensemble to get meaningful information on fluctuations. To this end, we introduce the Scaled Cumulant Generating Function (SCGF) 
\begin{equation}
    \label{eq:SCGF}
    \Psi_\eta(s) = \lim_{n \rightarrow \infty} \frac{1}{n} \ln G_{\eta,n}(s) = \lim_{n \rightarrow \infty} \frac{1}{n} \ln \mathbb{E} \left[ e^{n s T_n} \right] \ ,
\end{equation}
which characterizes the leading exponential behavior of the moment generating function $G_{\eta,n}(s)$ associated with $T_n$.
\newtext{Here, $s$, the Laplace parameter that enters in the SCGF, plays the role of the conjugate parameter to $T_n$. Intuitively, as much as the
inverse temperature in equilibrium statistical mechanics is connected to the internal energy
of a system through the derivative of the canonical free energy, $s$ is connected to the observable $T_n$. When $s >0 $, $T_n$ will more likely take values that are larger than the typical value
and viceversa when $s<0$. }
For finite and connected hypergraphs, $\Psi_\eta(s)$ is analytic, and one can calculate $I_\eta(t)$ via the Gartner--Ellis theorem~\cite{Ellis1985,DenHollander2000,Touchette2009,Dembo2010} that makes use of the Legendre--Fenchel (LF) transform
\begin{equation}
    \label{eq:LegendreFenchel}
    I_\eta(t) = \sup_{s \in \mathbb{R}} \left( s t - \Psi_\eta(s) \right) \ ,
\end{equation}
which links the Laplace parameter $s$ with a fluctuation $T_n = t$ as 
\begin{equation}
    \label{eq:LegendreDuality}
    t=\Psi_\eta'(s) \ .
\end{equation}
Because the random walk $X$ is ergodic, the SCGF can be obtained as
\begin{equation}
    \label{eq:SCGFSpectral}
    \Psi_\eta(s) = \ln \zeta_s \ ,
\end{equation}
where $\zeta_s$, computed numerically, is the dominant eigenvalue of the so-called tilted matrix 
\begin{equation}
    \Pi_s = \left\lbrace (\pi_s)_{ij} \right\rbrace = \left\lbrace \pi_{ij} e^{s \delta_{0,j}} \right\rbrace \ .
\end{equation}

%Results
%%Quenched average
To account for average properties of the ensemble of hypergraphs considered, one can take a quenched average over the disorder---here characterized by the number $\eta$ of higher-order interactions---of the function $\Psi_\eta$. Recalling that $H$ is a binomially distributed random variable with parameter $p$ and that the maximum number of higher-order interactions is $N_\triangle$, the quenched average can explicitly be written as
\begin{equation}
    \label{eq:QuenchedAverage}
    \Psi_{\text{q}}(s) = \sum_{\eta = 0}^{N_\triangle} \mathbb{P}(\eta) \Psi_\eta(s) \ ,
\end{equation}
where `$\text{q}$' stands for quenched

\footnote{Remarkably, the quenched average \eqref{eq:QuenchedAverage} takes such a simplified form because for a fixed number $\eta$ of higher-order interactions, we have only one possible transition matrix. However, we note that more complicated models might lead to different disorder configurations and therefore different transition matrices. In the latter case, to disentangle disorder and dynamics one would need to carefully study combinatorially how many different configurations arise by fixing $\eta$.}. Given $\Psi_{\text{q}}(s)$ in \eqref{eq:QuenchedAverage}, the quenched rate function $I_{\text{q}}(t)$ can be obtained via an LF transform of $\Psi_q$ (rather than $\Psi_\eta$) in \eqref{eq:LegendreFenchel}.

%%Effect of higher-order interactions on fluctuations

\begin{figure}
    \centering
    \includegraphics[width=\linewidth]{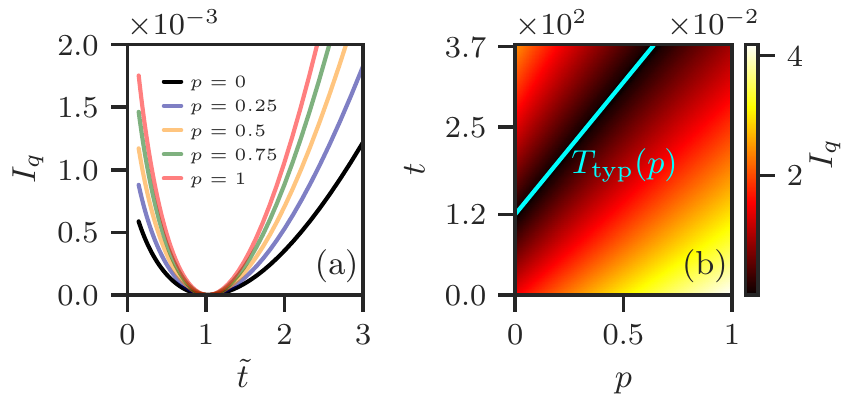}
    \caption{(a) Rate functions $I_q(\tilde{t})$ as a function of the rescaled time $\tilde{t}$ for different densities of higher-order interactions in the hypergraph $p$. The higher the $p$, the narrower the rate functions for $|\tilde{t}|>1$. (b) Heatmap representing how the rate function $I_q(t)$ behaves as a function of $t$ and $p$ (for visualisation purposes we plot $\sqrt{I_q}$). The light-blue line represents the typical value $T_{\text{typ}}$ which linearly increases with $p$. Plots obtained for a hypergraph with $N=1000$ nodes. %\leo{cambiare la figura con la nuova notazione $\hat{I}_q$}
    }
    \label{fig1}
\end{figure}

To understand the role of higher-order interactions, we first look at whether fluctuations of a given magnitude are more or less likely to appear on higher-order networks generated with different values of $p$. To understand this, we re-scale $t$ in $I_{\text{q}}(t)$ with the typical fraction of time spent in $0$ by the random walk at a fixed parameter $p$, namely $T_{\text{typ}}$, obtained by averaging \eqref{eq:TypicalTime} over $\mathbb{P}(\eta)$. In Fig.\ \ref{fig1}(a) we plot the rate functions $I_q(\tilde{t}=t/T_{\text{typ}})$ \newtext{($\tilde{t}$ is the time fraction on the core node relative to typical time)} for different values of $p$. %and compare them with the rate function for a graph with no higher-order interactions ($p=0$ case). 
Because of the re-scaling, all rate functions are $0$ at the typical value $\tilde{t}=1$. The likelihood is encoded in the shape of the rate function branches, the higher (lower) the branch the exponentially-less (more) likely is a fluctuation $\tilde{t} \neq 1$ to appear. 
We notice that with increasing $p$ the average number of higher-order interactions pointing to node $0$ grows generating a `confinement' effect,  \newtext{which has two consequences on the dynamics.}
%On the one hand
\newtexttwo{First}, at fixed $p$, fluctuations are more likely for times greater than the typical time, making it easier to visit the core node than peripheral nodes, as revealed by the asymmetric shape of the rate functions in Fig.\ \ref{fig1}(a). 
\newtexttwo{Moreover, as $p$ increases the transition towards the core node is favored, and fluctuations, both in excess and in deficit relative to the typical time, are hampered, as evidenced by the narrowing of the rate functions with increasing $p$ in Fig.\ \ref{fig1}(a).}

\newtexttwo{More in detail, in Fig.\ \ref{fig1}(b) we show how $I_q$ depends both on the non-rescaled time $t$ and $p$. We observe that the typical time increases with $p$  but also that relative time changes are associated with bigger absolute fluctuations (the level lines of $I_q(p,t)$ are not parallel to $T_{\text{typ}}$). }
Moreover, comparing with the case of a fully pairwise graph ($p=0$), on the one hand we show that the typical behavior at greater $p$ is atypical for the case $p=0$. On the other hand, rare values of $T_n$ greater than the typical one for the case $p=0$ can become typical just by increasing the number of higher-order interactions. By contrast, rare values of $T_n$ smaller than the typical one become even more atypical by introducing higher-order interactions.  %\st{Finally} 

%%Interaction of dynamics and structure

\vspace{0.5cm}

\paragraph{Annealed fluctuations}

We now consider random walks defined on non-static hypergraphs. Such annealed \cite{dorogovtsev2008critical} scenario is relevant to predict dynamical behaviors in time-varying systems where the structure evolves at a rate which is comparable to the time-scale of the process on top \cite{guerra2010annealed}, or in large systems whose precise characterization is often limited by lack of data or noise \cite{peralta2018stochastic}. 
In particular, we investigate the annealed fluctuations of the occupation time observable in \eqref{eq:Observable} over non-fixed realizations of three-body interactions for the model introduced above. In such a scenario, large fluctuations of a dynamical observable, such as $T_n$, could be generated by an optimal, albeit rare, realization of the underlying structure.

We consider the joint probability of obtaining a realization of the higher order structure and the occupation time in \eqref{eq:Observable}, and compute the moment generating function $G_{n}(s)$ associated with the observable $T_n$ with respect to this probability. We notice that $G_{n}(s)$ takes the form of an \textit{annealed} average of the moment generating function $G_{\eta,n}$ over the disorder

\begin{equation}
    \label{eq:AnnealedAverage}
    G_{n}(s) = \sum_{\eta=0}^{N_\triangle} \mathbb{P}(\eta) G_{\eta,n}(s) \ ,
\end{equation}
where we remind the reader that fixing $s$ corresponds to fixing a fluctuation $t$ (on average) according to \eqref{eq:LegendreDuality}.

We consider the regime of long times and large graphs, with the condition $n \gg N_\triangle \gg 1$, and introduce the fraction of total triangles $h = \eta/ N_\triangle$. The moment generating function $G_n(s)$ can be expressed using a saddle point approximation in $(h, t)$, i.e.,
\begin{equation}
    \label{eq:LimitAnnealedAverage}
    G_n(s) \approx e^{n \left( \ell^{-1} \log \mathbb{P}(h^*) + \Psi_{\eta^*}(s) \right)} \ ,
\end{equation}
where we call $\ell=n/N_\triangle$ the \emph{annealing parameter} and indicate the saddle-point solution with $(h^*, t^*)$, adopting the shorthand notation $\eta^* = h^* N_\triangle$. In the following, we focus on the non-trivial exponent of \eqref{eq:LimitAnnealedAverage}:
\begin{equation}
    \label{eq:LeadingBehaviour}
    \hat{\Psi}_\ell(s) \coloneqq \ell^{-1} \log \mathbb{P}(h^*) + \Psi_{\eta^*}(s) \ .
\end{equation}

We can obtain the annealed SCGF from \eqref{eq:LeadingBehaviour} by taking the infinite $\ell$ limit, that is $\Psi_a(s) \coloneqq \hat{\Psi}_{\ell \rightarrow \infty}(s)$. The function $\Psi_a(s)$, together with its LF transform $I_a(t)$, completely describes atypical fluctuations of occupation times in the annealed regime. For large values of $\ell$, disorder and dynamics `interact' at the saddle-point solution of \eqref{eq:LimitAnnealedAverage} selecting the most likely structure that realises the occupation-time fluctuation associated with $s$.  We remark that \eqref{eq:LimitAnnealedAverage} is valid as long as $\ell$ is large \footnote{In particular, for $\ell$ finite and small, one has $N_{\triangle} > n$ and therefore the ergodicity assumption necessary to derive $\Psi_{\eta^*}(s)$ falls.}. 

However, since the disorder is self-averaging, in the limit $\ell \rightarrow 0$ all probability concentrates around the typical number of higher-order interactions, recovering the quenched average \eqref{eq:QuenchedAverage} for a fixed $p$.

\begin{figure}
    \centering
    \includegraphics[width=\linewidth]{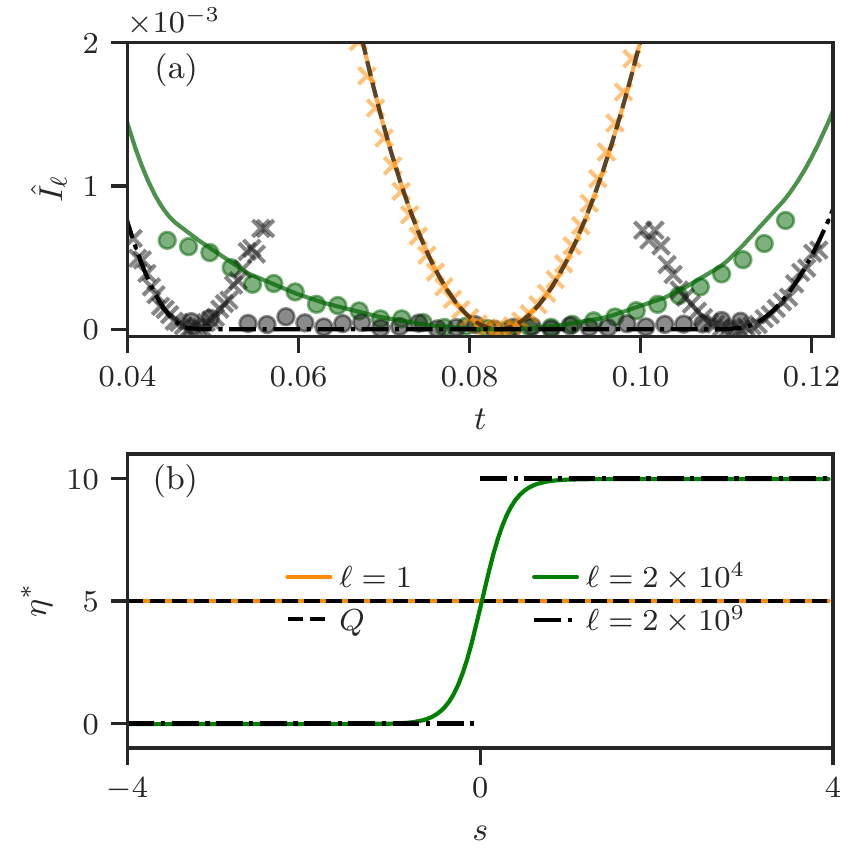}
    \caption{\newtext{(a) Rate functions $\hat{I}_\ell$ functions for different $\ell$, as a function of $t$. Monte-Carlo quenched simulations for the three cases with no (left-most), maximum (right-most), and half-maximum (center)  number of higher-order interactions are plotted as cross-shaped scatter points. Annealed simulations' results are plotted as round scatter points for different values of $\ell$, colored according to the legend. (b) The optimal value $\eta^*$ of the number of higher-order interactions plotted as a function of $s$ (the fluctuation parameter).   Results are obtained for a hypergraph with $N=21$ nodes and $p=0.5$.}}
    \label{fig2}
\end{figure}

%%%%HERE
%%Explain figure, end with ergodicity breaking
In Fig.\ \ref{fig2}(a) we plot $\hat{I}_\ell$ for several values of $\ell$. As expected, for small $\ell$ we retrieve the quenched rate function $I_q$ (for the parameter $p=0.5$ used here) which is realised by the typical number of higher-order interactions $\eta^* = h^* N_\triangle \sim \text{ceil} \left[ N_\triangle/2 \right]$ throughout all fluctuations shown in Fig.\ \ref{fig2}(b). As we increase $\ell$, the function $\hat{I}_\ell$ tends to flatten, and in the limit $\ell \rightarrow \infty$ the annealed rate function $I_a$ develops a plateau of zeros \footnote{That an annealed rate function is a lower bound of a quenched one is known in the mathematics literature~\cite{Greven1994,Comets2000,Varadhan2003,Zeitouni2006}. Intuitively, this is consequence of picking an optimal structure to generate fluctuations in the dynamics rather than having it fixed as in the quenched case.}.
Although $I_a$ exhibits a continuous range where it equals zero, not every occupation time $t$ within this range is a typical event. \newtexttwo{Within the saddle-point approximation in \eqref{eq:LimitAnnealedAverage}, it appears that o}nly the times resulting from the most probable network configurations, which manifest at the boundaries of this zero plateau, truly represent the typical behavior of the observable $T_n$.
These specific configurations, \newtexttwo{as shown in Fig.\ \ref{fig2}(b)}, are statistically favored and dominate the ensemble. 
To further validate our observations, in Fig.\ \ref{fig2}(a) we also plot Monte-Carlo (MC) simulations for both the quenched and annealed case. \newtexttwo{Details on how to perform such simulations and their physical interpretations are reported in \textit{Appendix B} and \textit{C}}. Quenched simulations appear as coloured cross-shaped scatter points for three different scenarios of random walks exploring a graph with no (left-most gray), max (right-most gray), and half-max (\newtext{orange}) number of higher-order interactions. Annealed simulations appear as enlarged green and gray scatter points for two different values of $\ell$. 
In particular, orange %\st{point} 
crosses well describe the shape of the quenched rate function $I_q$ and gray circles well show the flattening of the function $\hat{I}_\ell$ at large values of $\ell$.   
\newtext{Noticeably, from the saddle-point calculation in Fig.\ \ref{fig2}(b) it is evident that for large values of $\ell$} as one slightly moves from the typical scenario $s=0$ and looks into fluctuations for either $s<0$ or $s>0$, the structure $\eta^*$ optimally realising such fluctuations abruptly changes from, respectively, a graph with no higher-order interactions, i.e., $\min{(\eta^*)}=0$, to a structure that maximizes their number, i.e., $\max{(\eta^*)}=10$ for $N = 21$. 

For finite $\ell$ we observe a continuous crossover centered in $s=0$ between these two regimes. For large $\ell$, such crossover appears to be much steeper, hinting at the existence of a transition in the limit $\ell\rightarrow \infty$ between two regimes, one where the random walk spreads over the entire graph, and one where it spends more time on the core node due to higher-order interactions. \newtexttwo{As discussed in \textit{Appendix C} and in the SM, this behavior is an artifact of the saddle-point approximation. Indeed,} the existence of a phase transition is not confirmed by an analysis of the distribution of $T_n$ at large $\ell$ of simulations of random walks on evolving hypergraphs, which converge to an unimodal distribution with $\ell \rightarrow \infty$. This suggests that the observed flattening might be due to neglecting sub-leading $o(n)$ terms in Eq.\ \eqref{eq:LimitAnnealedAverage}.
\newtexttwo{Furthermore, the lack of an exponential scaling in the bulk distribution of $T_n$ indicates that typical fluctuations} occur more frequently.
In summary, while the saddle-point solution 
is limited in describing fluctuations of the system close to the typical time, it allows to correctly capture the extreme values of the annealed rate function, as confirmed by the good matching between MC simulations and analytical predictions in the tails of the rate functions.

\vspace{0.5cm}

\paragraph{Conclusion}

In this work we have shed light on the impact of higher-order interactions on the atypical behaviors of dynamical processes on networks. 
In particular, we have investigated random walks dynamics in a simplified higher-order model, a fully connected pairwise graph with additional random three-body interactions connecting a core node with peripheral nodes.
%, which allowed us to gain insights on dynamical fluctuations of diffusive processes in hypergraphs. 
By applying large deviations tools we have derived the leading exponential scaling of fluctuations for a dynamical observable, here considered to be the mean fraction of time the random walk spends on the system nodes. We characterized the dynamics of the system in two different scenarios, showing that the presence of higher-order interactions greatly affects rare events and atypical dynamics. In the quenched case, where the structure of the system is fixed, higher-order interactions inhibit random walk fluctuations of the occupation time at the core.
\newtexttwo{Conversely, in the SM, we show that fluctuations of the occupation time on peripheral nodes are enhanced far off the typical occupation time.}
In the annealed case, averaging over dynamics on non-fixed structures, the random walk dynamics select the optimal structure that realises a particular fluctuation. In such a scenario, \newtexttwo{fluctuations} of the occupation time are more likely to appear, and by means of a saddle-point approximation, it is possible to capture dynamical fluctuations far from the typical time. \newtexttwo{In the SM, we validated our results on complex structures and showed that homogeneous hypergraphs exhibit a non-trivial density of higher-order interactions boosting fluctuations.} Finally, results shown here for random walks extend to broader dynamics, such as for large values of the biasing parameter for biased random walks on hypergraphs, where the bias promotes or hampers the visit of nodes with many higher-order interactions. In the future, it might be interesting to broaden our understanding of the impact of specific higher-order structural features, such as %hypergraphs displaying a 
scale-free distribution of higher-order interactions~\cite{kovalenko2021growing}, community structure~\cite{contisciani2022inference}, or directed hyperedges~\cite{gallo1993directed}.

\newtexttwo{Eventually}, our work might be proven useful \newtexttwo{also} to characterize the appearance of rare and catastrophic events in the interconnected structure of higher-order systems, or to control patterns of infections in adoption and rumour diffusion in real-world social networks.

% \vspace{0.5cm}

 \paragraph{Acknowledgments}
L.D.G. acknowledges Paolo D. Piana and Francesco D. Ventura for the fruitful discussions. L.D.G. thanks O. Sadekar for the help.

\vspace{0.5cm}

\appendix
\setcounter{equation}{0}

\renewcommand{\theequation}{A.\arabic{equation}}

\setcounter{figure}{0}
\renewcommand{\thefigure}{A.\arabic{figure}}

\paragraph{Appendix A - Transition matrix of random walk on hypergraphs}
In the random walk on hypergraph the walker chooses with equal probability among its hyperlinks and then selects one of the nodes belonging to such a higher-order structure, favouring intrinsically those neighbours that belong to highest-order hyperlinks. 
In order to write the transition matrix, we start defining the hyper incidence matrix $e_{i\alpha}$ telling if a node $i$ belong to a hyperlink $E_\alpha$, namely:
\begin{equation}
e_{i\alpha} =
\begin{cases}
1 & \text{if } i \in E_{\alpha} \\
0 & \text{otherwise}
\end{cases}.
\end{equation}

From the hyperincidence matrix one can define the hyperadjacency matrix as follows:
\begin{equation}
    A = ee^T,
\end{equation}
where $A_{ij}$ represents the number of hyperlinks containing both nodes $i$ and $j$.
Furthermore, one can build the hyperedges matrix, $C_{\alpha \beta}$,
\begin{equation}
     C = e^T e, 
\end{equation}
 whose entry $C_{\alpha \beta}$ counts the number of common nodes between $E_{\alpha}$ and $E_{\beta}$ ($E_{\alpha} \cap E_{\beta} $) and $C_{\alpha \alpha}$ is the size of an hyperlink $E_\alpha$, or equivalently its order of interaction plus one, $|E_\alpha| = O_\alpha +1 $.

By means of $C$ and $e$, we can construct the weight of the transition matrix of the unbiased random walk, $ k_{ij}^H $, that reads,
 
\begin{equation}
    k_{ij}^H = \sum_\alpha (C_{\alpha \alpha}-1 )e_{i\alpha}e_{j\alpha} =(e\hat{C}e^T)_{ij}- A_{ij}, 
    \label{weight_trans_matrix}
\end{equation}
where its entries represent the sum of the orders of all the common hyperlinks between $i$ and $j$. 
Summing $k_{ij}^H $ over all neighbours of a node $i$, one obtains the order-weighted hyperdegree, 
\begin{equation}
    k_{i}^H = \sum_l k_{il}^H,
    \label{order-weighted degree}
\end{equation}

namely the sum of the orders of all the hyperlinks belonging to $i$.

Therefore, the transition matrix of the unbiased random walk on a hypergraph reads 
\begin{equation}
\Pi_{ij}= \frac{\sum_\alpha (C_{\alpha \alpha}-1 )e_{i\alpha}e_{j\alpha}}{\sum_l \sum_\alpha (C_{\alpha \alpha}-1 )e_{i\alpha}e_{l\alpha}} =  \frac{ k_{ij}^H}{\sum_l  k_{il}^H } = \frac{ k_{ij}^H}{ k_{i}^H }.
\label{trans_matrix_1}
\end{equation}

\paragraph{Appendix B - Quenched Monte-Carlo simulations}

Given a hypergraph of size $N$ with a configuration of higher-order interactions $\eta$ sampled from the binomial distribution in Eq.\ \eqref{eq:probability_triangles}, we run simulations of length $n$. The result of this is a histogram of values for the observable $T_n$  for a given hypergraph. We then calculate the rate function (see Eq.\ \eqref{eq:LegendreFenchel}) for the observable $T_n$ as 
\begin{equation}
    I_{\eta}^{\text{sim}}(t) = - \frac{1}{n} \ln \mathbb{P}^{\text{hist}}_{\eta}(t) \, ,
\end{equation}
where superscript `$\text{sim}$' indicates that the function is obtained from `simulations' and `$\text{hist}$' refers to the fact that the distribution is approximated by the `histogram' related to the simulations. We repeat the procedure for many configurations of the hypergraph randomly selected from the binomial distribution in Eq.\ \eqref{eq:probability_triangles}  and calculate the rate functions by averaging as follows
\begin{equation}
    I_{\text{q}}^{\text{sim}}(t) = \sum_{\eta=0}^{N_\triangle} \mathbb{P}^{\text{hist}}(\eta) I_{\eta}^{\text{sim}}(t) \, ,
\end{equation}
where $\mathbb{P}^{\text{hist}}(\eta)$ is the probability distribution of configurations $\eta$ at a fixed $p$ obtained with the random generation of graphs (it converges to Eq.\ \eqref{eq:probability_triangles} ).
Notice that the cumulative statistics over different hypergraphs come only after re-scaling with $1/n \ln$ each distribution of $T_n$. 
These are the quenched simulations represented as gray ($p=0$ and $p=1$) and orange ($p=0.5$) circular dots in Fig.\ \ref{fig2} (a).  They are used as a sanity check both for the quenched limit of our annealed calculation for $p=0.5$ in the middle and, in the case of the annealed rate function, to check that the extrema of the zeros plateau corresponds to the two opposite situations of a graph with no triangular interactions for $p=0$ (on the left) and a graph with $N_\triangle$ (the maximum possible) triangular interactions for $p=1$ (on the right).

\paragraph{Appendix C - Annealed Monte-Carlo simulations}
In order to carefully calculate  the Legendre transform of Eq.\ \eqref{eq:LeadingBehaviour}, which is the asymptotic leading behaviour of Eq.\ \eqref{eq:AnnealedAverage}, and visualise the rate functions appearing in Fig. \ref{fig2}(a) we generate many trajectories of the random walk of length $n$ (which in turn fixes the parameter $\ell = n/N_\triangle$ for a graph of $N$ nodes) where each one is initialised over a hypergraph with a number of triangular interactions picked up at random from the binomial distribution in Eq.\ \eqref{eq:probability_triangles}. The graph is resampled over the trajectory of the random walk at a fast rate. 
Once all the trajectories are obtained we calculate the cumulative statistic (the histogram) of the observable $T_n$ and, only after that, re-scale the properly normalised histogram by $1/n \ln$. It is important to stress here that in the annealed scenario the re-scaling comes after obtaining the full statistics over all hypergraphs for the observable $T_n$ (notice that this procedure is inverted in the quenched scenario), which is the reason why at the saddle point of Eq.\ \eqref{eq:LimitAnnealedAverage} dynamics and disorder `interact'. This procedure already generates a distribution $\mathbb{P}^{\text{hist}}_{\text{a}}$ for the observable $T_n$ and from it we directly calculate the rate function
\begin{equation}
    I_{\text{a}}^{\text{sim}}(t) = - \frac{1}{n} \ln \mathbb{P}^{\text{hist}}_{\text{a}}(t) \, .
\end{equation}

This is the procedure followed to obtain the annealed simulations plotted in Fig.\ref{fig2} (a). 

\begin{figure}[h]
    \centering
    \includegraphics[width=\linewidth]{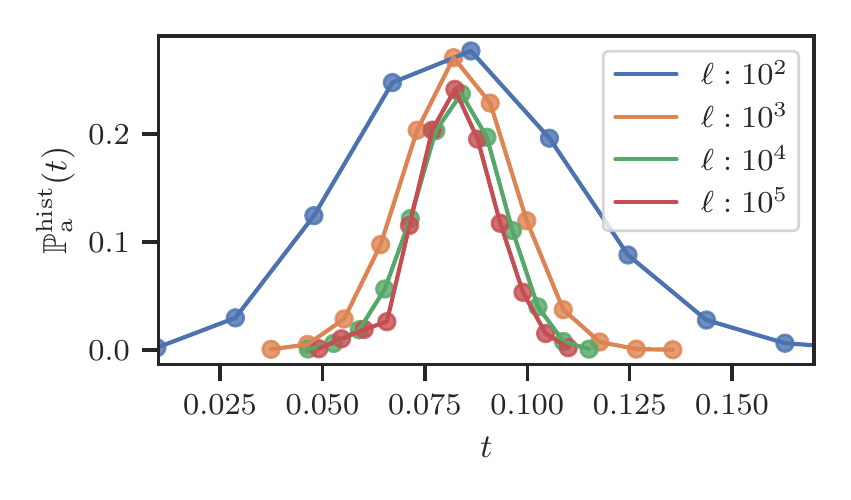}
    \caption{Histograms of observable $T_n$ from annealed simulations at different values of $n$ ($\ell=n$). The simulations are performed considering $N=21$, and $10^5$ different trajectories.}
    \label{fig:histogram}
\end{figure}

The histograms of $\mathbb{P}^{\text{hist}}_{\text{a}}$ for different values of $n$ reveal that there is no observable flattening across the simulations. Instead, as $n$ increases, the histograms converge, indicating no true phase transition in the system, see Fig. \ref{fig:histogram}. This suggests that the flattening of the rate function observed in the annealed scenario is caused by solely examining the saddle point in the study of dynamics using large deviations, neglecting sub-leading contributions.

\bibliography{mybib}

\newpage
\onecolumngrid
\vspace{\columnsep}

\large
\section*{Dynamical fluctuations of random walks in higher-order networks: Supplemental material}

\normalfont
%\tableofcontents
\renewcommand{\thefigure}{S\arabic{figure}}
\setcounter{figure}{0}

\setcounter{equation}{0}

\renewcommand{\theequation}{S.\arabic{equation}}

{
In the main paper, we have analyzed the behavior of dynamical fluctuations in a simple higher-order network with a star-like structure \newtext{in a quenched and annealed scenario, where the structure of interactions is either fixed or evolving in time}.
\newtext{
In this Supplemental Material (SM) we further validate our results with a series of additional analysis, and discussing simulations on more complex topologies and alternative and more general classes of dynamical processes. In particular:

\begin{itemize}
    \item In I we give details on how to derive the transition matrix of a random walk over a general higher-order network.

    \item In II we extend our study for unbiased random walks to biased random walks analyzing both quenched and annealed fluctuations.

    \item In III we discuss quenched dynamical fluctuations on peripheral nodes of the star-like structure investigated in the main text.

    \item In IV we give more details on the flattening of the rate function observed in the annealed scenario using  large-deviation tools.

    \item In V we extend the study by analyzing atypical dynamical behavior in three additional star-like hypergraph models with more complex topology.

    \item In VI we investigate fluctuations in a different type of hypergraph model, with no preferential node, referred to as the homogenous hypergraph model.

    \item In VII we discuss in detail how to run quenched and annealed simulations and show histograms of the latter.
\end{itemize}
}}

\renewcommand{\thesection}{S\arabic{section}}
\setcounter{section}{0}

\newtext{
\section{I Transition matrix of unbiased random walks on higher-order networks}
\label{SecI}

In this Section, we provide a detailed characterization of the random walk on hypergraph, introduced in \cite{carletti2020random}. In a first-order unbiased random walk, a walker in a node $i$ moves to one of its neighbours $j$ choosing with equal probability among its links. In the case of a higher-order unbiased random walk, we want to define a dynamics in which the walker chooses with equal probability among its hyperlinks and then selects one of the nodes belonging to such a higher-order structure, favouring intrinsically those neighbours that belong to highest-order hyperlinks. Such a dynamic portrays the diffusion of ideas or the spreading of rumours or misinformation that is generally enhanced by group interactions.
In order to write the transition matrix of this process, we start defining the hyper incidence matrix $e_{i\alpha}$ telling if a node $i$ belong to a hyperlink $E_\alpha$, namely:
\begin{equation}
e_{i\alpha} =
\begin{cases}
1 & \text{if } i \in E_{\alpha} \\
0 & \text{otherwise}
\end{cases}.
\end{equation}

From the hyperincidence matrix one can define the hyperadjacency matrix as follows:
\begin{equation}
    A = ee^T,
\end{equation}
where $A_{ij}$ represents the number of hyperlinks containing both nodes $i$ and $j$.

Furthermore, one can build the hyperedges matrix, $C_{\alpha \beta}$,
\begin{equation}
     C = e^T e, 
\end{equation}
 whose entry $C_{\alpha \beta}$ counts the number of common nodes between $E_{\alpha}$ and $E_{\beta}$ ($E_{\alpha} \cap E_{\beta} $) and $C_{\alpha \alpha}$ is the size of an hyperlink $E_\alpha$, or equivalently its order of interaction plus one, $|E_\alpha| = O_\alpha +1 $.

By means of $C$ and $e$, we can construct the weight of the transition matrix of the unbiased random walk, $ k_{ij}^H $, that reads,
 
\begin{equation}
    k_{ij}^H = \sum_\alpha (C_{\alpha \alpha}-1 )e_{i\alpha}e_{j\alpha} =(e\hat{C}e^T)_{ij}- A_{ij}, 
    \label{weight_trans_matrix SM}
\end{equation}
where its entries represent the sum of the orders of all the common hyperlinks between $i$ and $j$. For instance if two nodes $i$ and $j$ share one link, two second-order hyperlinks (three body interactions) and one third-order hyperlink (four body interactions), $k_{ij}^H = 1\times 1+ 2\times 2 + 1 \times 3 = 8$.

Summing $k_{ij}^H $ over all neighbours of a node $i$, one obtains the order-weighted hyperdegree, 
\begin{equation}
    k_{i}^H = \sum_l k_{il}^H,
    \label{order-weighted degree SM}
\end{equation}

namely the sum of the orders of all the hyperlinks belonging to $i$.

Therefore, the transition matrix of the unbiased random walk on hypergraphs reads 
\begin{equation}
\Pi_{ij}= \frac{\sum_\alpha (C_{\alpha \alpha}-1 )e_{i\alpha}e_{j\alpha}}{\sum_l \sum_\alpha (C_{\alpha \alpha}-1 )e_{i\alpha}e_{l\alpha}} =  \frac{ k_{ij}^H}{\sum_l  k_{il}^H } = \frac{ k_{ij}^H}{ k_{i}^H }.
\label{trans_matrix_1 SM}
\end{equation}

Note that in the case of simple graphs, having only first-order interactions, and therefore $C_{\alpha \alpha} = 2$ for every link $E_\alpha $, we obtain the transition matrix of the unbiased random walk on simple graphs:
\begin{equation}
\Pi_{ij}= \frac{\sum_\alpha (C_{\alpha \alpha}-1 )e_{i\alpha}e_{j\alpha}}{\sum_l \sum_\alpha (C_{\alpha \alpha}-1 )e_{i\alpha}e_{l\alpha}} =   \frac{2\sum_\alpha e_{i\alpha}e_{j\alpha}- A_{ij}}{2\sum_l \sum_\alpha e_{i\alpha}e_{l\alpha}-k_i} = \frac{A_{ij}}{k_i}.
\end{equation}

% \section{General class of random walks on higher-order networks}

%\textcolor{blue}{Definire classe generale di hyperedegree-biased random walk (ispirati da Gardenes Latora PRE 2009). Riprodurre Fig.2a for a fixed p ma differenti alpha, ad esempio 0, -1, -2, 1 e 2. Mostrare anche una figura al variare di alpha come variabile continua sulla x. Scrivere questa sezione in blu. Aggiungere link a questa sezione nel main.}

% We can extend the random walk dynamic defined in the latter section introducing a more general class of dynamical processes on hypergraphs that depend only on the sizes of hyperlinks, $C_{\alpha \alpha}$.
% Specifically, we assume that the transition probability of a random walker on a hypergraph depends on a generic function of the size of the hyperlinks $\varphi({C_{\alpha \alpha}})$ \cite{carletti2020dynamical}.
% We can generalize Eq. \ref{trans_matrix_1} as follows,
% \begin{equation}
% \Pi_{ij}^\varphi = \frac{\sum_\alpha \varphi(C_{\alpha \alpha})e_{i\alpha}e_{j\alpha} }{\sum_l \sum_\alpha \varphi(C_{\alpha \alpha})e_{i\alpha}e_{j\alpha} }.
% \label{phi_trans_matrix}
% \end{equation}

% $\Pi_{ij}^\varphi$, is the $\varphi$-dependent transition probability, whose elements indicate that the probability to jump from $i$ to $j$ is proportional to $\varphi(C_{\alpha \alpha})$.

% Notice that, we retrieve the unbiased random walk case by choosing as $\varphi$ the order of the interactions (size of the hyperlink minus one):
% \begin{equation}
% \varphi(C_{\alpha\alpha}) = C_{\alpha\alpha} -1 = O_\alpha.
% \end{equation}

\section{II Biased random walks on higher-order networks}

\label{SecII}

Here we introduce a new class of random walks on hypergraphs, specifically a hyperdegree-biased random walk. In analogy with the biased random walk on simple graphs \cite{gomez2008entropy}, such a dynamic extends the unbiased random walk introducing a bias that enhances or hampers the attractiveness of nodes with respect to their hyperdegree. 

For the first-order case (pairwise interactions only), every node $j$ has a bias equal to $k_j^\gamma$ ($k_j$ is the degree of $j$) and the transition matrix reads
\begin{equation}
    \Pi_{ij}^B =\frac{A_{ij}k_j^\gamma}{\sum_j A_{ij}k_j^\gamma} ,
\end{equation}
where $\gamma$ is the bias exponent. For $\gamma >0$, the transition towards large-degree nodes is favoured, while for $ \gamma <0$ nodes with small degrees attract the walker more. For $\gamma= 0 $, the transition matrix retrieves the unbiased case, being $k_j^\gamma=1$.

Following a similar approach, for the higher-order case we can define a hyperdegree-biased random walk that depends on the order-weighted hyperdegree  $k_j^H$.
Resorting the definition of $k_j^H$ from Eq.\ \eqref{weight_trans_matrix} and \eqref{order-weighted degree}, we can write the transition matrix as

\begin{equation}
\Pi_{ij}^{HB} =\frac{k_{ij}^H (k_j^H)^\gamma}{\sum_l k_{il}^H (k_l^H)^\gamma}.
\end{equation}

Again, this dynamic favours the transition towards nodes with large $k^H$ when $\gamma >0$, makes less attractive the same nodes when $\gamma <0$ and returns the unbiased case shown in Eq.\ \eqref{trans_matrix_1} for $\gamma = 0$.

Intuitively, by means of this generalization of the random walk on hypergraph, we can portray a large class of processes where one might need to tune the effects of group interactions. For instance, the diffusion of trends or norm adoption can be accelerated in large groups because of conformism mechanisms and peer pressure ($\gamma >0$), or exploratory behaviors in information-seeking processes, where individuals or algorithms prioritize novel or less popular sources over well-known ones ($\gamma <0$).

\subsection*{Results}
In this Subsection, we detail the results of the biased random walk, previously defined. As an observable, we focus on the fraction of time spent on the core node over the hypergraph model introduced in the main material. Fig.\ \ref{fig:Biased_1} shows our findings for the quenched calculations. In panel (a), we plot the rate function for diverse values of the bias parameter $\gamma$. We observe that the rate functions (as a function of the rescaled time $\tilde{t}$) are narrower at increasing $\gamma > 0$, indicating that an attractive bias towards higher-order interactions reduces fluctuations beyond the unbiased scenario. This aligns with our observations in the main text where higher-order interactions were seen to suppress fluctuations, with increased attractiveness further intensifying the `confinement' effect and thus, reducing fluctuations. In contrast, $\gamma < 0$ leads to a broader rate function, signifying that this kind of higher-order interactions allows for larger fluctuations.

Panel (b) of Fig.\ 1 provides a complementary analysis by displaying the rate function in relation to both $\gamma$ and $t$. %This examination elucidates the combined impact of bias and time on dynamical fluctuations, reinforcing our understanding of how fluctuation dynamics are influenced by these parameters.

\begin{figure}[h]
    \centering    \includegraphics{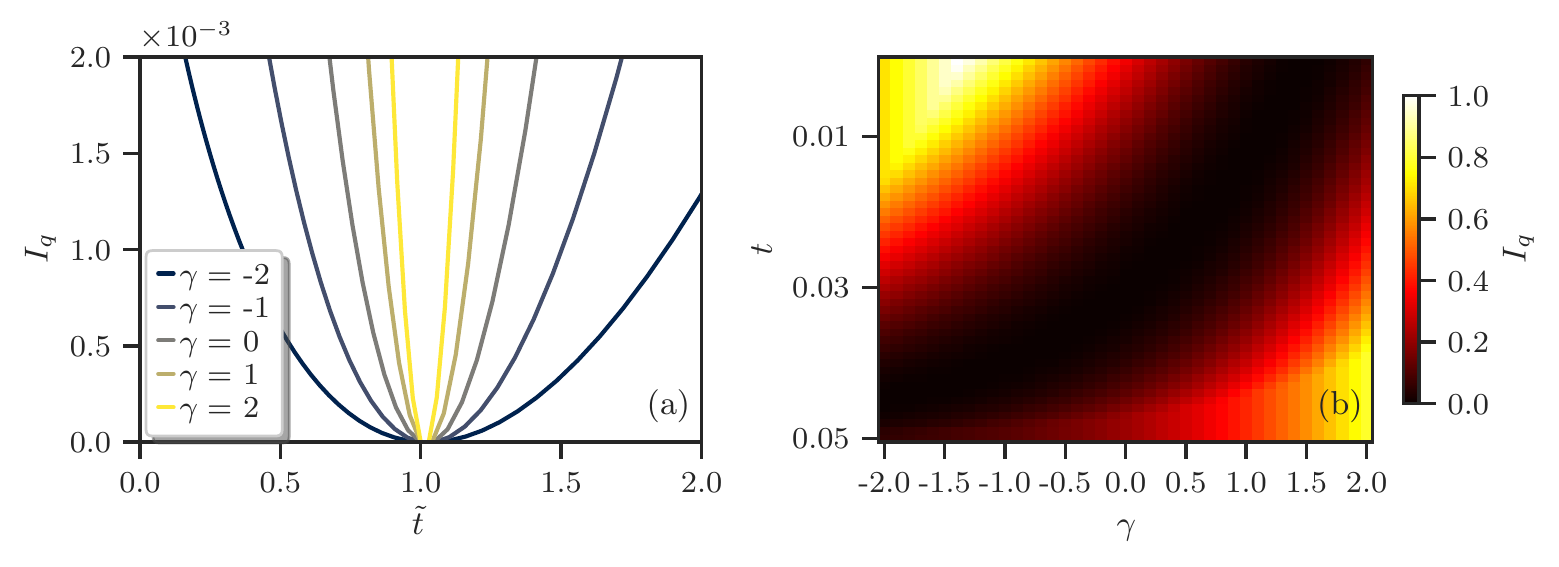}
    \caption{(a) Rate functions $I_q(\tilde{t})$ as a function of the rescaled time $\tilde{t}$ for different bias parameters $\gamma$. The larger the $\gamma$, the narrower the rate functions for $|\tilde{t}|>1$. (b) Heatmap representing how the rate function $I_q(t)$ behaves as a function of $t$ and $\gamma$.  Plots obtained for a hypergraph with $N=100$ nodes and density of higher-order interactions $p = 0.5$.}
    \label{fig:Biased_1}
\end{figure}

In Fig.\ \ref{fig:Biased_2} we show two heatmaps for distinct $\gamma$ values, displaying the rate function's dependency on $t$ and $p$, akin to the approach in Fig.\ 2(b) in the main text. %These heatmaps allow for a detailed comparison of how the rate function varies across temporal and parameter spaces under specific bias settings, offering insights into the complex dynamics at play.

\begin{figure}[h]
    \centering    \includegraphics{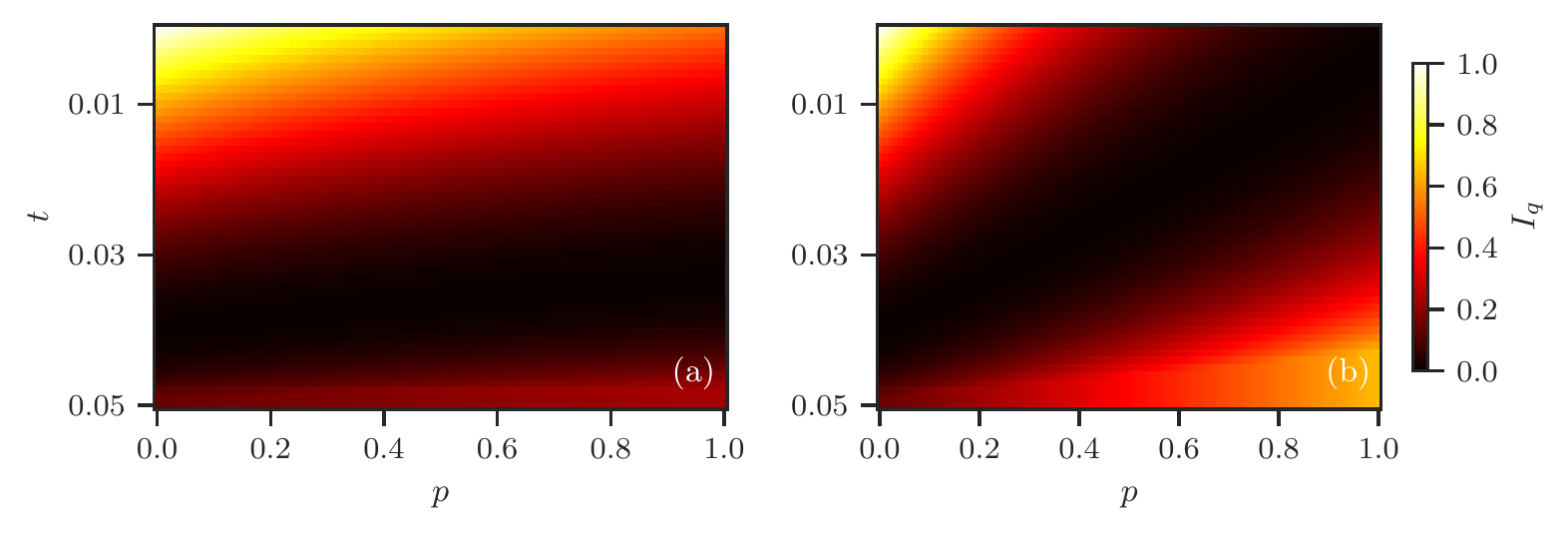}
    \caption{Heatmaps representing how the rate function $I_q$ behaves as a function of $t$ and $p$ for two different values of $\gamma$: $\gamma = -0.5$ panel (a) and $\gamma= 0.5$ panel (b).}
    \label{fig:Biased_2}
\end{figure}

% Through this exploration, we aim to provide a comprehensive understanding of the biased random walk's dynamical fluctuations within the hypergraph model, emphasizing the significant role of the bias parameter in modulating these fluctuations.

% \textcolor{red}{write that the results are obatained for the model of the illustration}

Turning our attention to the annealed case, Fig.\ \ref{fig:rate annealed biased} delineates the rate functions for varying annealed parameters across two distinct values of $\gamma$. Remarkably, in Fig. \ref{fig:rate annealed biased}, we observe a flattening of the rate function equivalently to the unbiased case of Main Text
(see discussion on the limits of the validity of the saddle-point approach presented in the main text and in the SM to fully capture such a behavior). 

% This phase transition, regardless of the value of $\gamma$, underscores the intrinsic nature of the dynamical process and its inherent sensitivity to the system's higher-order interaction framework. The alignment of the rate function plateaus with specific interaction regimes--—namely, the absence or full presence of higher-order interactions--—highlights the critical role these interactions play in dictating the system's dynamical behavior.
%The plateaus observed in the rate functions align with the typical times when either no higher-order interactions are present or they are fully engaged.

\begin{figure}[h!]
    \centering
    \includegraphics[width = \textwidth]{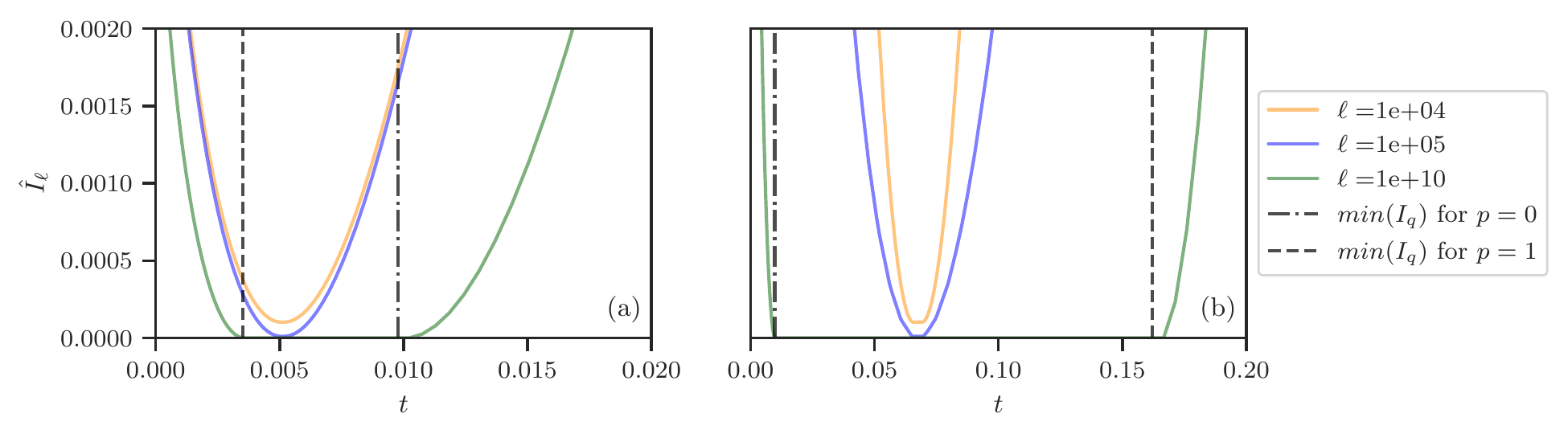}
    \caption{Annealed rate function $\hat{I_\ell}$  for different $\ell$ as a function of $t$, and the characteristic times (vertical dashed lines) for $\eta = 0$ and $\eta=max(\eta)$. The results are obtained considering $\gamma = -2$ (a) and $\gamma = 2$ (b).
    }
    \label{fig:rate annealed biased}
\end{figure}

\begin{figure}[h!]
    \centering
    \includegraphics[width = \textwidth]{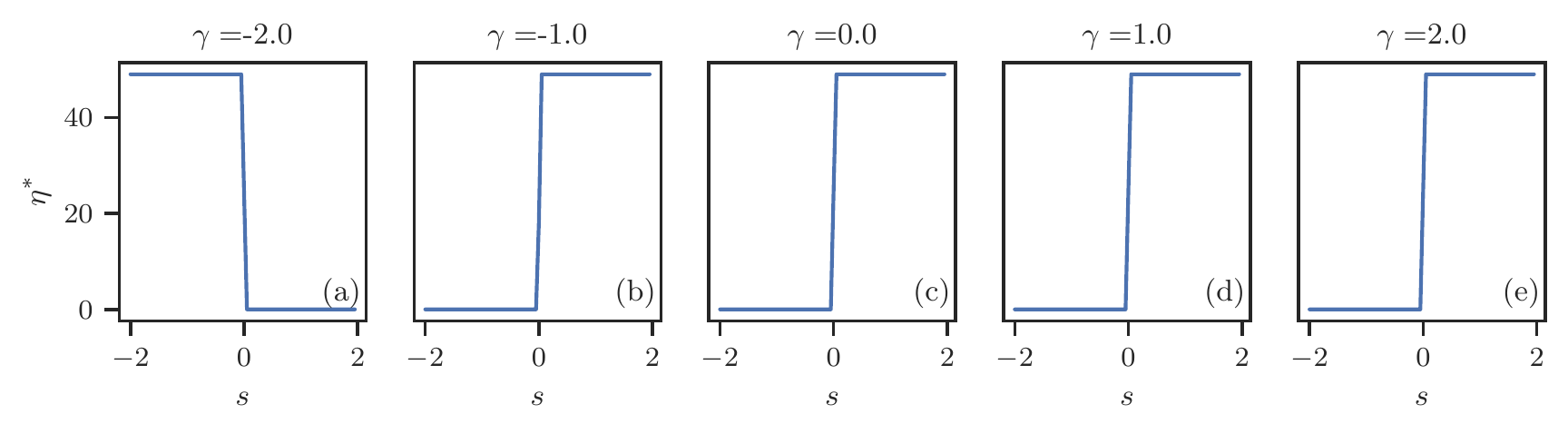}
    \caption{The optimal value $\eta^*$ of the number of higher-order interactions as a function of the tilting parameter $s$ for $p=0.5$, and different $\gamma: [-2,-1,0,1,2]$.}
    \label{fig:eta annealed biased}
\end{figure}

Lastly, we examine which configurations of higher-order interactions maximize the fluctuations in the annealed scenario considering different values of the bias parameter $\gamma$. In fig. \ref{fig:eta annealed biased}, we show that the biased random walks maximize fluctuations in two different regimes. For $\gamma > -1$, the dynamical system behaves accordingly to the unbiased case, $\gamma = 0$, having the optimal configuration with no higher-order interactions for fluctuations of the residence time on the core node smaller than the typical value ($\eta^* =0$, for $s<0$) and with the totality of such interactions for fluctuations of the residence time greater than the typical value ($\eta^* =N_\Delta$, for $s>0$).
On the contrary, for $\gamma < -1$ in Fig.\ \ref{fig:eta annealed biased} (a), the optimal configurations are inverted with respect to positive and negative fluctuations.
Intuitively, in the original unbiased case, the higher-order interactions increase the transition probability proportionally to the generalized hyperdegree $k^H_i$, as discussed in Section \textit{I} of this Supplemental Material.
When one inserts a negative bias with respect to $k^H_i$, the attractivity of nodes with higher-order interactions is reduced, and at $\gamma = -1$ the topological bias on higher-order interactions is compensated dynamically, with this interplay between structure and dynamics making the random walk not feel the effect of higher-order interactions anymore. 
\begin{figure}[h!]
    \centering
    \includegraphics[width = \textwidth]{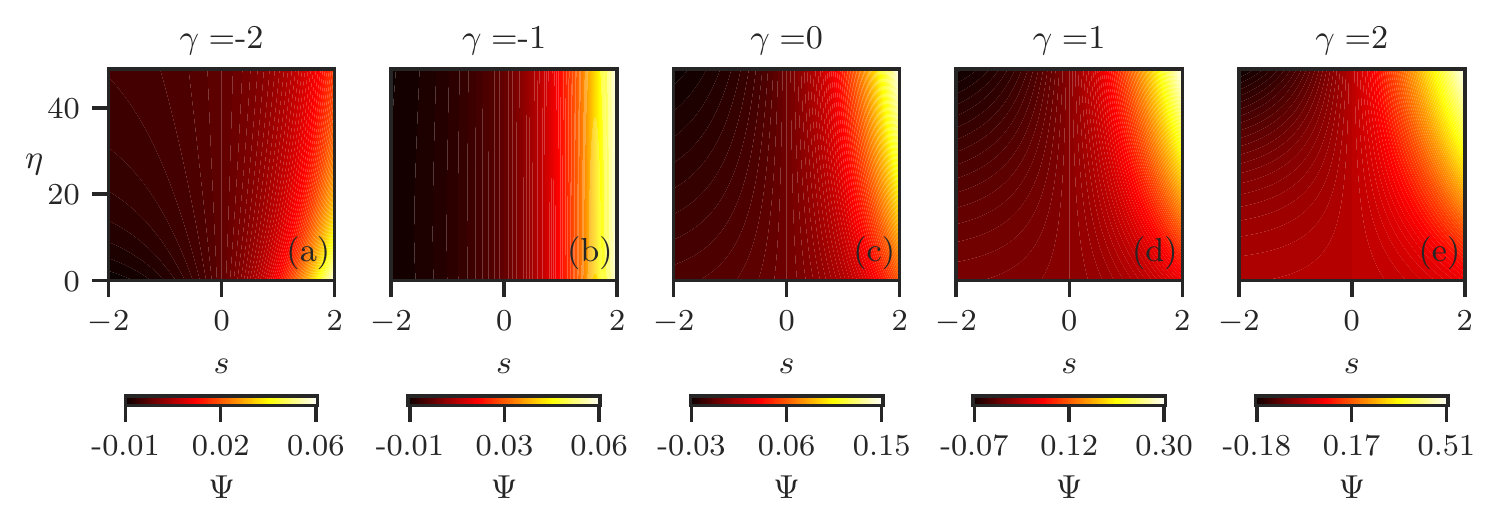}
    \caption{Heatmaps representing how the SCGF $\Psi$ behaves as a function of $\eta$ and $s$ for five different values of $\gamma: [-2,-1, 0,1,2]$. }
    \label{fig: heat map annealed biased}
\end{figure}

To further validate this idea, we plot in Fig. \ref{fig: heat map annealed biased} the SCGF $\Psi$ with respect to the number of triangular interactions $\eta$ and the tilting parameter $s$.
We observe that, for $\gamma=-1$, $\Psi$ does not depend on $\eta$ (panel (b)) and therefore is independent with respect to the number of higher-order interactions. While for $\gamma <-1$ (panel (a)), $\Psi$ has an inverted behaviour with respect to the case $\gamma > -1$ (panels (c), (d), (e)).
}

\newtext{
\section{III Dynamical fluctuations on peripheral nodes}
\label{SecIII}

In this section we investigate the dynamical fluctuations of the time a random walk spends on  peripheral nodes of the model presented in the main text. Specifically, in Fig.\ \ref{fig:periferical nodes}, we show the rate functions considering as observable the occupation time on peripheral nodes (all nodes but the core), i.e., $\bar{T}_n = \frac{1}{n} \sum_{l=1}^n \sum_{i=1}^{N-1} \delta_{X_l,i}$. We refer to $\bar{I}_q(\tilde{t})$ as the rescaled rate function associated with the new observable $\bar{T}_n$. Noticeably, we observe the opposite behavior, with an enhancement of fluctuations far off the typical occupation time. In summary, by introducing higher-order interactions on a fully-pairwise network we make it easier for the random walk to spend more (less) time on the core (peripheral) node(s).

\begin{figure}[h!]
    \centering
    \includegraphics{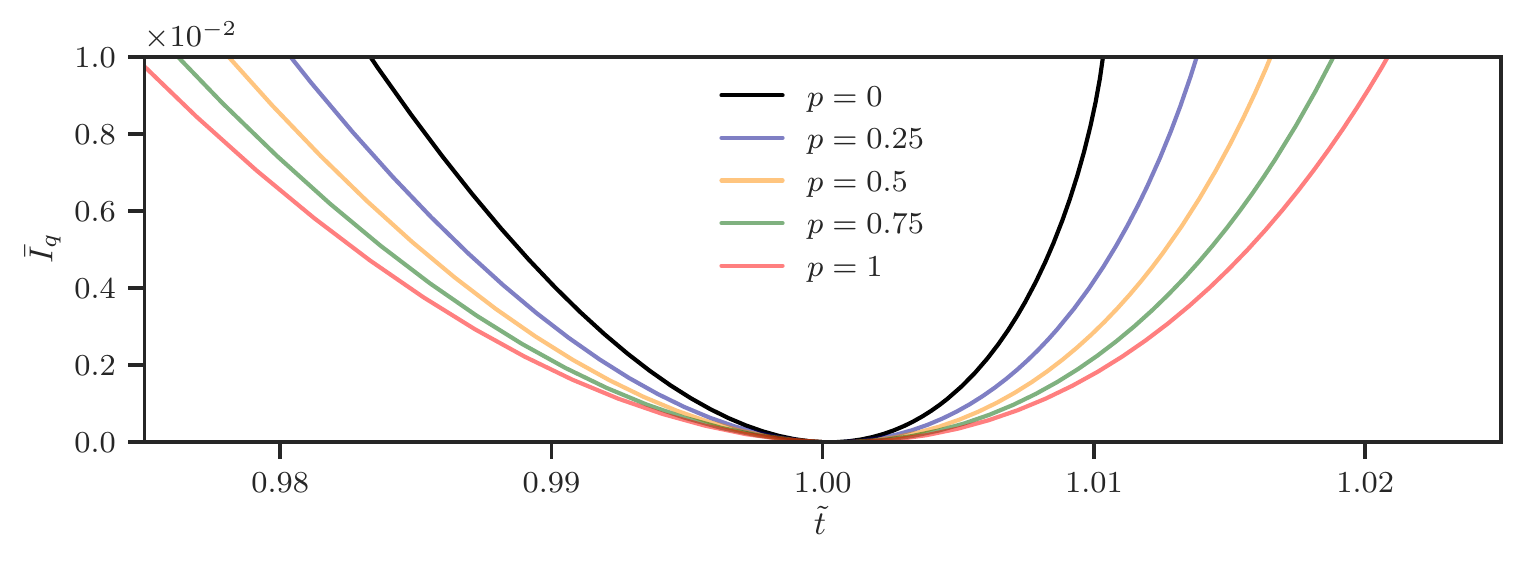}
    \caption{ Rate functions $\bar{I}_q(\tilde{t})$ as a function of $\tilde{t}$ observing the occupation time on all peripheral nodes ($\{1, \ldots N-1 \}$).}
    \label{fig:periferical nodes}
\end{figure}}

\newtext{
\section{IV Flattening of the rate function in the annealed scenario}
\label{SecIV}

In this Section, we further investigate the flattening of the rate function derived from large-deviation theory tools in the annealed scenario and presented in the main text. In Section \ref{SecVII}, after having defined how to numerically calculate rate functions in the annealed scenario, we will give further insights on the nature of such a presumed phase transition, showing that what we observe is actually caused by solely examining the saddle point in the study of
dynamics neglecting sub-leading contributions.

Given this, resorting to the saddle point approximation defined in Eq. (13) of the Main material, as presented in Fig.\ 3, we observe that when one considers the two regimes of optimal number of higher-order interactions $\eta^*$ for which the system maximizes the fluctuations, (one for $\eta^* =0$ for $s<0$ and another one for $\eta^* =max(eta)$ for $s>0$),
for finite $\ell$ we observe a continuous crossover centred in $s=0$ between these two regimes. For large $\ell$, such crossover appears to be much steeper, suggesting the existence of a phase transition in the limit $\ell\rightarrow \infty$. However, as discussed in detail in Section
\ref{SecVII}, where the large deviation solution is compared with simulations of random walk on evolving hypergraphs, this is due to neglecting
 sub-leading terms in the saddle point approximation.

In addition, we show the scaling analysis in  Fig.\ \ref{fig:phase_transition_plot} (b) where we observe a power-law decay of $\Delta_s = s_{\max{(\eta^*)}} - s_{\min{(\eta^*)}}$, i.e., the distance in terms of $s$ between the two extreme three-body interaction regimes, as a function of $\ell$.
Lastly, in Fig.\ \ref{fig:phase_transition_plot}(a), we plot $\Psi_q(s)$ and $\Psi_a(s)$. For the latter, differently from the quenched case, we observe a discontinuity in the first derivative at $s=0$.

\begin{figure}[h]
    \centering
    \includegraphics[width = \linewidth]{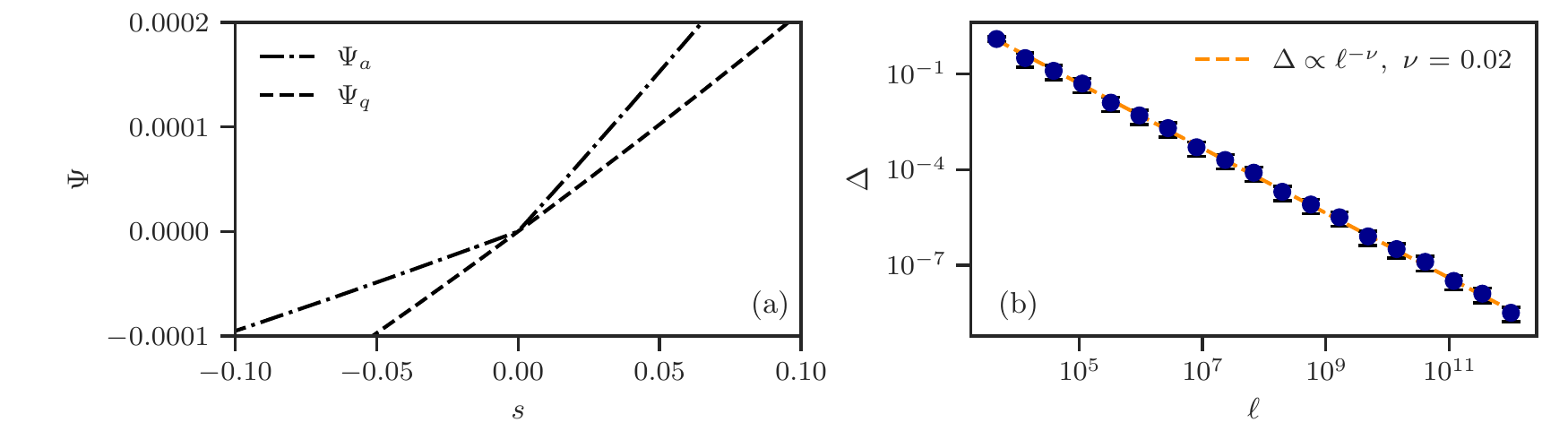}
    \caption{(a) Quenched $\Psi_q$ and annealed $\Psi_a$ SCGFs as a function of $s$. The latter shows a discontinuity in its first derivative at $s=0$. (b) Scaling, in log-log scale, of the width \( \Delta_s \) as a function of the annealing parameter \( \ell \).}
    \label{fig:phase_transition_plot}
\end{figure}
}
%\st{Since the SCGF $\Psi_a(s)$ is considered to be the free energy of the system}~\cite{Touchette2009} \st{the appearance of a discontinuity in its first derivative suggests the existence of a first-order phase transition in the limit $\ell \rightarrow \infty$.}

\section{V Dynamical fluctuations in star-like hypergraph models with more complex topology}
\label{SecV}

% Redefine figure numbering and reset figure counter

{
In this section we discuss the atypical behavior of random walks on three additional star-like hypergraph models. While these models preserve the main features behind the model considered in the main text, their analysis presents additional complications. The main text model has the advantage of being extremely simple from a combinatorial point of view, allowing us to compute both quenched and annealed averages without making use of a numerical sample of all possible realizations of the model. Specifically, Eq.(1) describes the probability of drawing a hypergraph with a certain number of three-body interactions, and inserted in Eqs. (11) and (14) allows us to obtain respectively the quenched and the annealed Scaled Cumulant Generating Function. By contrast, these additional models are more costly because the weights $\mathbb{P}(\eta) $ used in both the quenched and annealed average (Eq. (11) and (12)) can only be found by an extensive numerical sample.}

\subsubsection*{Model S1: Core-node, All-possible triangles, underlying complete pairwise graph}
{
Model S1 extends Model A by allowing the formation of all possible triangles that include the central node $0$ and any two peripheral nodes \( i \) and \( j \). These triangles are generated with a probability \( p \). Similar to Model A, this model also features a complete pairwise graph that fully interconnects all nodes.}

\subsubsection*{Model S2: Core-node, Non-overlapping triangles, underlying random regular pairwise graph}
{
Model S2 maintains the same central node and non-overlapping triangles as the original Model A. However, the underlying pairwise connectivity is described by a regular random graph with pairwise degree $k=3$, rather than being a complete graph. This introduces additional randomness in the connectivity patterns of the nodes.}

\subsubsection*{Model S3: Core-Node, All-possible triangles, underlying random regular pairwise graph}
{
Model S3 combines elements of Model S1 and S2. In particular, it includes a central node $0$ and allows the formation of all possible triangles involving \( 0 \) and any pair of peripheral nodes \( i \) and \( j \) as in Model S1. These triangles are realized with a probability \( p \). The underlying pairwise connectivity of this model is a regular random graph with pairwise degree $k=3$. }

\subsection*{Results}
\begin{figure}[h]
    \centering
    \includegraphics[width = \textwidth]{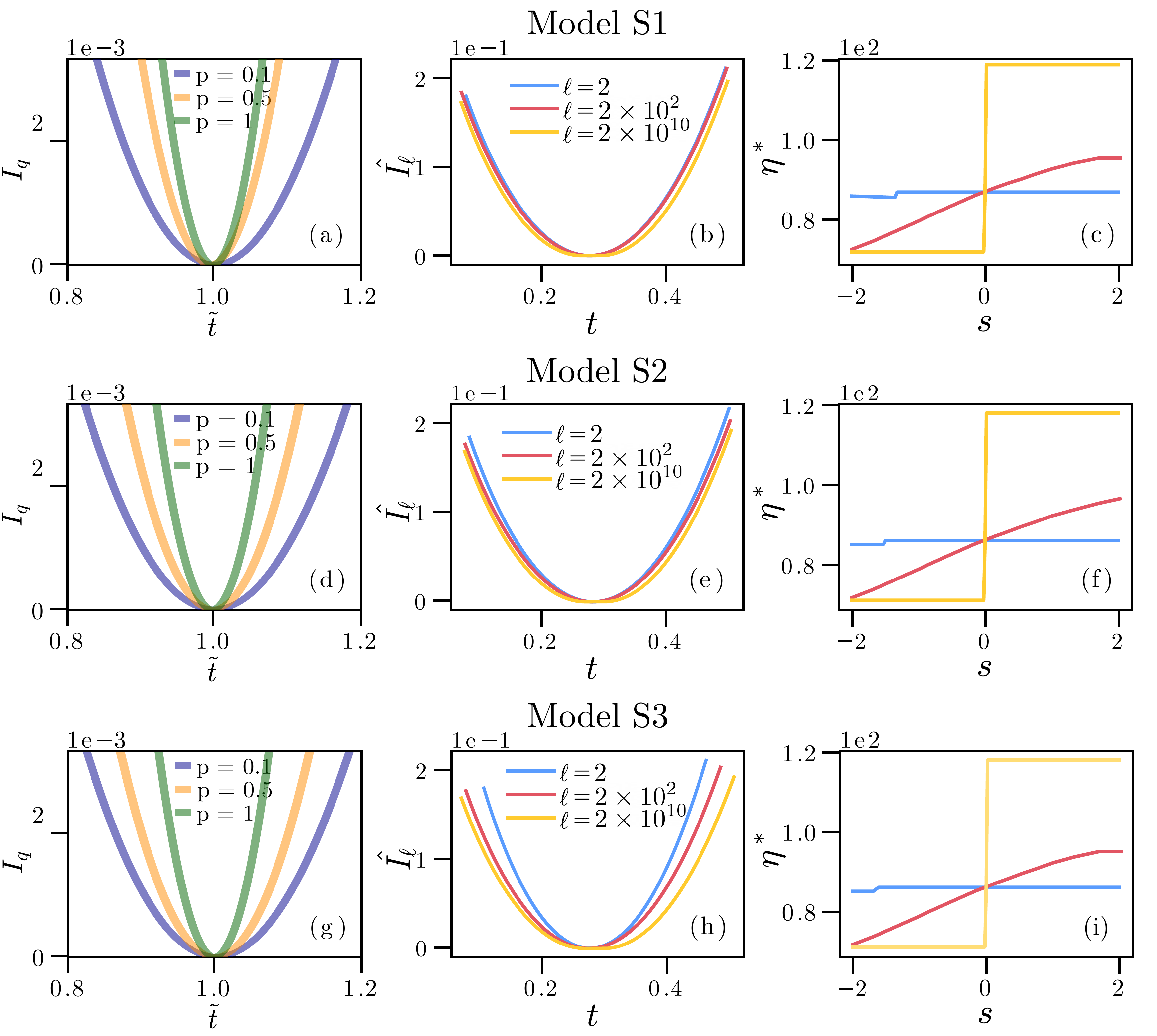}
    \caption{ { \small (a,d,g) Quenched rate functions $I_q(\tilde{t})$ for different densities of higher-order interactions $p$. (b,e,h) Functions $\hat{I}_\ell(t)$ for different values of the annealing parameter $\ell$ for $p=0.5$. (c,f,i) The optimal value $\eta^*$ for the number of higher-order interactions as a function of the tilting parameter $s$ for $p=0.5$. Results are obtained for hypergraphs generated for Models S1 (top row), S2 (middle) and S3 (bottom), with $N=20$ nodes.}}
    \label{fig:SM 1}
\end{figure}

{
In Fig.\ \ref{fig:SM 1} we investigate dynamical fluctuations in both the quenched and annealing scenario by reproducing some of the plots presented in the main text for Models S1 (top row), S2 (middle row) and S3 (bottom row). In particular, the left panels (a, d and g) refer to the quenched scenario discussed in Figure 2(a) of the main text. By contrast, the middle (b, e and h) and right panels (c, f and i) characterize the annealed scenario and should be compared with Figure 3(a) and Figure 3(b) respectively.}
{
In particular, in Fig.\ \ref{fig:SM 1}, panels (a, d and g) for the quenched scenario we plot the quenched rate functions $I_q(\tilde{t}=t/T_{\text{typ}})$ as a function of the rescaled time $\tilde{t} = t/T_{\text{typ}}$ for various values of $p$. The rate function reflects the likelihood of dynamical fluctuations, with a higher one indicating a lower probability for fluctuations with $\tilde{t} \neq 1$. We observe that, for $p>0$, the presence of higher-order interactions consistently reduces the probability of deviations from the typical value, thereby restricting the random walk's ability to visit either a core-localized or periphery-delocalized phase during fluctuations. Increasing $p$ the average number of higher-order interactions pointing to node $0$ grows generating the same `confinement' effect on the dynamics discussed in the main paper. As a consequence, escaping from node $0$ becomes harder and dynamical fluctuations are suppressed.}
{
In Fig.\ \ref{fig:SM 1}, panels (b, e and h), we display the annealed rate functions $\hat{I}_\ell$ sa a function of the rescaled time $\tilde{t}$ for various levels of the annealing parameter $\ell$, for a density of higher-order interactions given by $p=0.5$. 
For lower values of $\ell$, we recover the quenched rate function $I_q$, which is realised by the typical number of higher-order interactions $\eta^*$ across all fluctuations. For higher values of $\ell$, the function $\hat{I}_\ell$ begins to exhibit a flattening trend, in analogy to what we observed in the simpler model discussed in the main text.}
{
Lastly, in Fig.\ \ref{fig:SM 1}, panels (c,f and i), for the same three values of $\ell$ we plot the optimal number of three-body interactions $\eta^*$ that corresponds to the minimum of the rate functions in panels (b, e and h), namely the specific configuration that maximizes the fluctuations over the annealed average.
For the lowest value of $\ell$, the behavior of $\eta^*$ is practically undistinguishable from the quenched scenario. For an intermediate value of the annealing parameter, $\ell=2\times10^2$, we observe that for negative values of $s$ the optimal fluctuations are obtained for small $\eta^*$ , while for positive values of $s$ these are obtained for large $\eta^*$. The continuous crossover between these two regimes, centered in $s=0$, becomes much steeper for the highest value of the annealing parameter, $\ell=2\times 10^9$, supporting the existence of a transition between such two regimes in the limit of $\ell\rightarrow \infty$.}
{
In summary, the patterns of dynamical fluctuations observed in the more complex higher-order topologies considered in Models S1, S2 and S3, are consistent to the ones observed in the simpler model of the main text, for both the quenched and annealed scenario. }

%The plots confirm and prove the solidity of our main findings about dynamical fluctuations both on the quenched and annealed scenarios. 

\clearpage
\newpage

\section{VI Dynamical fluctuations in a homogeneous higher-order network}
\label{SecVI}

{
In the following, we investigate a model with no preferential / core node, where the three-body interactions are randomly distributed among any triplet of nodes $(i,j,k)$ with probability $p$ \newtext{on top of a fully-connected structure} in the quenched scenario.}

\begin{figure}[h!]
    \centering
    \includegraphics[width=.7\linewidth]{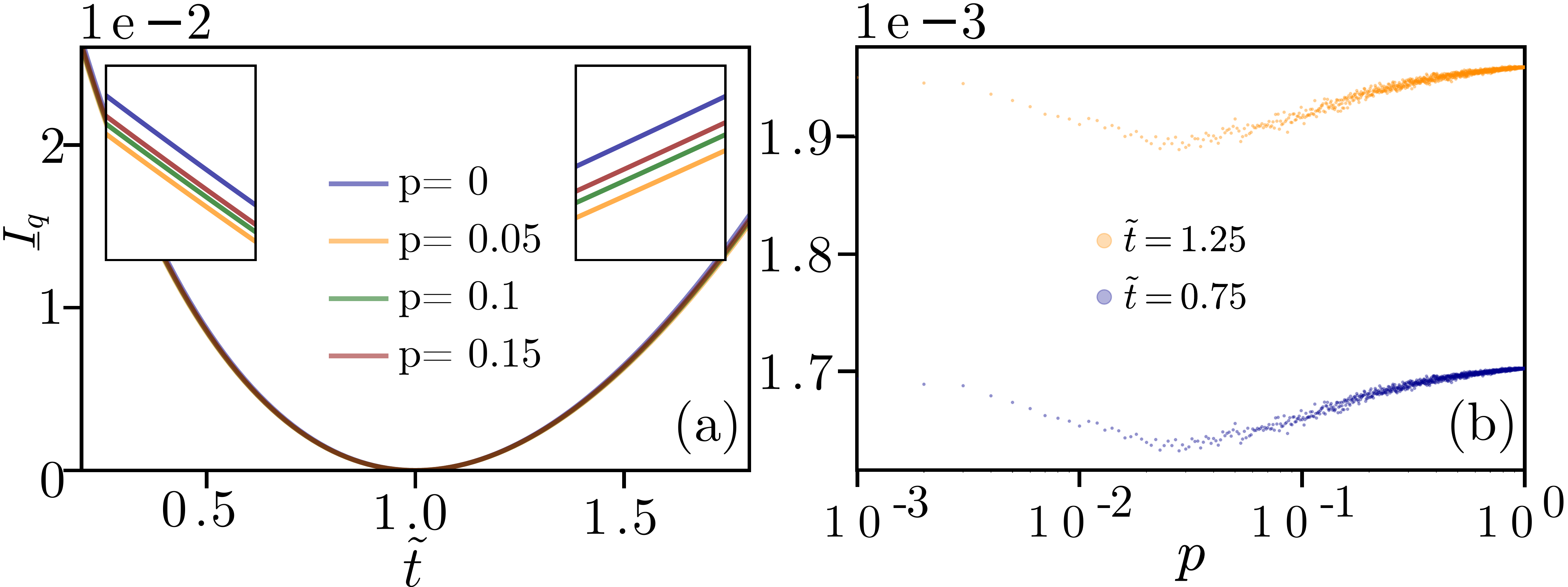}
    \caption{ { \small (a) Rate function $\underline{I}_q(\tilde{t})$ as a function of the rescaled time $\tilde{t}$ for several densities of three-body interactions $p$ in a hypergraph of $N=20$ nodes with homogeneously distributed triangles. (b) $\underline{I}_q$ for two fluctuations $\tilde{t}$ as a function of $p$ for two values of $\tilde{t}$.}}
    \label{fig1_appendix}
\end{figure}

{
In particular, in Fig. \ref{fig1_appendix}(a), we plot the large deviation rate function \( \underline{I}_q(\tilde{t}) \)  associated with the occupation-time observable $\underline{T}_n = \frac{1}{n} \sum_{l=1}^n \delta_{X_l,j}$ for a randomly chosen node $j$ as a function of the rescaled time $\tilde{t}$ for several values of $p$. 
The two insets zoom on the non-monotonic tails --with respect to $p$-- of the rate functions for values of $\tilde{t}$ far from $T_{typ}$.
Additionally, in Fig. \ref{fig1_appendix}(b), we plot the rate function, \( \underline{I}_q(\tilde{t}) \) as a function of $p$, for two values of $\tilde{t}$, one larger and one smaller than the typical time $T_{typ}$.}

{
In summary, the magnitude of fluctuations in homogeneous hypergraphs with no preferential core node display a non-monotonic dependence on the density $p$ of higher-order interactions. Remarkably, this indicates the existence of an optimal value of $p$ that minimizes $\underline{I}_q(\tilde{t})$, and therefore maximizes the appearance of atypical occupation times.}

\newtext{
\section{VII Monte Carlo simulations for the quenched and annealed scenarios}
\label{SecVII}

\subsection{Quenched simulations}

Given a hypergraph of size $N=21$ with a configuration of higher-order interactions $\eta$ sampled from the binomial distribution in Eq.\ (1) of the main text, we run $10^4$ simulations (the more the smoother the statistic) of length $n=10^4$ (the number of time steps of the random walk). The result of this is a histogram of values for the observable $T_n$ (fraction of time the random walk has spent on the core node) for a given hypergraph. We then calculate the rate function (see Eq.\ (7)) for the observable $T_n$ as 
\begin{equation}
    I_{\eta}^{\text{sim}}(t) = - \frac{1}{n} \ln \mathbb{P}^{\text{hist}}_{\eta}(t) \, ,
\end{equation}
where superscript `$\text{sim}$' indicates that the function is obtained from `simulations' and `$\text{hist}$' refers to the fact that the distribution is approximated by the `histogram' related to the simulations. We repeat the procedure for $10^4$ configurations of the hypergraph randomly selected from the binomial distribution in Eq.\ (1) of the main text and calculate the rate functions by averaging (similarly to how we average the SCGFs in Eq.\ (11) of the main text) as follows
\begin{equation}
\label{eq:QuenchedAverage SM}
    I_{\text{q}}^{\text{sim}}(t) = \sum_{\eta=0}^{N_\triangle} \mathbb{P}^{\text{hist}}(\eta) I_{\eta}^{\text{sim}}(t) \, ,
\end{equation}
where $\mathbb{P}^{\text{hist}}(\eta)$ is the probability distribution of configurations $\eta$ at a fixed $p$ obtained with the random generation of graphs (it converges to Eq.\ (1) of the main text for infinitely many simulations).
Notice that the cumulative statistics over different hypergraphs come only after re-scaling with $1/n \ln$ each distribution of $T_n$. 

These are the quenched simulations represented as gray ($p=0$ and $p=1$) and orange ($p=0.5$) circular dots in Fig.\ 3(a) of the main text.  They are used as a sanity check both for the quenched limit of our annealed calculation for $p=0.5$ in the middle and, in the case of the annealed rate function, to check that the extrema of the zeros plateau corresponds to the two opposite situations of a graph with no triangular interactions for $p=0$ (on the left) and a graph with $N_\triangle$ (the maximum possible) triangular interactions for $p=1$ (on the right).

\subsection{Annealed simulations}

In order to carefully calculate (from simulations) the Legendre transform of Eq.\ (14), which is the asymptotic leading behaviour of Eq.\ (12), and visualise the rate functions appearing in Fig.\ 3(a) of the main text we generate $10^5$ trajectories (the more the smoother the statistics) of the random walk of length $n=10^4$ (which in turn fixes the parameter $\ell = n/N_\triangle$ for a graph of $N=21$ nodes) where each one is initialised over a hypergraph with a number of triangular interactions picked up at random from the binomial distribution in Eq.\ (1). The graph is resampled over the trajectory of the random walk at a fast rate. For the simulations shown the graph is resampled at every time step of the random walk. However, we have seen that changing the rate slightly does not qualitatively change the results. 

Once all the trajectories are obtained we calculate the cumulative statistic (the histogram) of the observable $T_n$ and, only after that, re-scale the properly normalised histogram by $1/n \ln$. It is important to stress here that in the annealed scenario the re-scaling comes after obtaining the full statistics over all hypergraphs for the observable $T_n$ (notice that this procedure is inverted in the quenched scenario), which is the reason why at the saddle point of Eq.\ (13) in the main text dynamics and disorder `interact'. This procedure already generates a distribution $\mathbb{P}^{\text{hist}}_{\text{a}}$ for the observable $T_n$ and from it we directly calculate the rate function
\begin{equation}
    I_{\text{a}}^{\text{sim}}(t) = - \frac{1}{n} \ln \mathbb{P}^{\text{hist}}_{\text{a}}(t) \, .
\end{equation}

This is the procedure followed to obtain the annealed simulations plotted in Fig.\ 3 (a). These, as expected, show a flattening of their shape towards the asymptotic annealed behaviour at increasing $\ell$ confirming our annealed large deviation approach to study fluctuations of the observable $T_n$.
}

\subsubsection*{Histograms of $T_n$ in annealed simulations for different $n$}
\newtext{
In this section, we present the histogram of the observable $T_n$ obtained from annealed simulations for different values of $n$. For all simulations, we consider the annealing parameter $\ell = n$, where $n$ represents the time length of the simulations, with each simulation performed as described in the previous section. 
The histograms reveal a significant finding: there is no observable flattening across the simulations. Instead, as $n$ increases, the histograms converge, indicating no true phase transition in the system. This suggests that the flattening of the rate function observed in the annealed scenario is caused by solely examining the saddle point in the study of dynamics using large deviations, neglecting sub-leading contributions. 

\begin{figure}[h]
    \centering    \includegraphics[width = 16 cm]{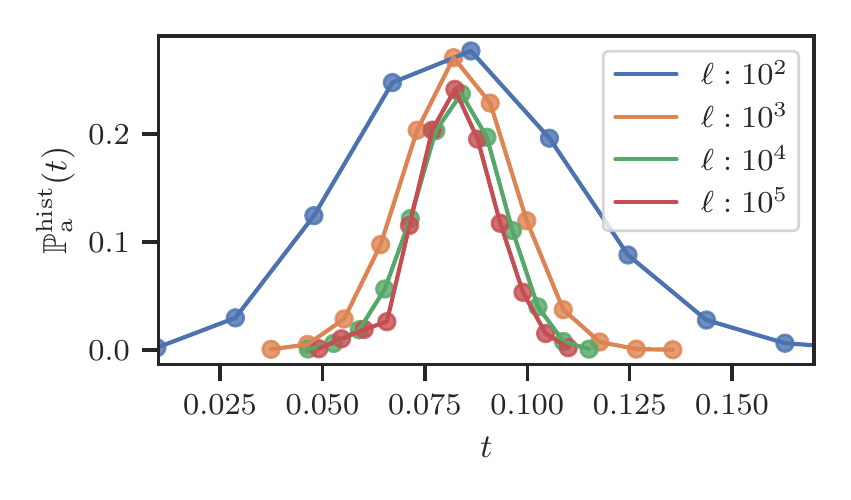}
    \caption{Histograms of observable $T_n$ from annealed simulations at different values of $n$ ($\ell=n$). The simulations are performed considering $N=21$, and $10^5$ different trajectories.}
    \label{fig:hist sm}
\end{figure}
}

\end{document}